\documentstyle[amssymb,preprint,aps]{revtex}

\begin{document}
\draft
\title{Chan-Paton Soliton Gauge States of Compatified Open String}
\author{Jen-Chi Lee\thanks{
e-mail: jclee@cc.nctu.edu.tw}}
\address{Department of Electrophysics, National Chiao-Tung University, Hsinchu,\\
30050,Taiwan}
\maketitle

\begin{abstract}
We study the mechanism of enhanced gauge symmetry of bosonic open string
compatified on torus by analyzing the zero-norm soliton (nonzero winding of
wilson line) gauge states in the spectrum. Unlike the closed string case, we
find that the soliton gauge state exists only at massive levels. These
soliton gauge states correspond to the existence of enhanced massive gauge
symmetries with transformation parameters containing both Einstein and
Yang-Mills indices. In the T-dual picture, these symmetries exist only at
some discrete values of compatified radii when N D-branes are coincident.
\end{abstract}


\section{Introduction}

The discovery of D-brane as R-R charge carrier\cite{1} and its applications
to various string dualities has made it clear that open string is essential
in the study of string theory. Historically, Yang-Mills gauge symmetry was
incorporated into string theory through different mechanisms for closed and
open strings. For the closed string, it was built into the theory through
compatification of string coordinates or, more generally, by adding an
internal Kac-Moody conformal field theory with the appropriate central
charge \cite{2}. For the open string instead, Yang-Mills degree of freedom
was built into the theory through Chan-Paton effect\cite{3} by adding
charges at the end points of string.

In the previous paper\cite{4}, we related the closed string Kac-Moody gauge
symmetry to the existence of massless {\em zero-norm} soliton gauge states
(SGS) in the spectrum of torus compatification. The program was then
extended to the massive states. The existence of massive SGS thus implies
that there is an infinite enhanced gauge symmetry of compatified closed
string theory. In this paper, we will study the SGS of compatified {\em open}
string theory. Unlike the closed string case, we find that the SGS exists
only at massive levels. These SGS correspond to the enhanced massive
symmetries with transformation parameters containing both Einstein and
Yang-Mills indices. This is reminiscent of the symmetry of closed Heterotic
string massive modes discovered previously\cite{5}. In the T-dual picture,
these SGS implies the existence of enhanced massive gauge symmetry at some 
{\em discrete} values of compatified radii when N D-branes are coincident.

This paper is organized as following. In section II we discuss the
uncompatified open string. We first derive both massless and massive
Chan-Paton zero-norm gauge states. The corresponding gauge symmetries and
Ward identities are then derived. In the massive case, we get mixed
Einstein-Yang-Mills-type symmetry, which is similar to the one we derived in
the closed Heterotic string theory. Section III is devoted to the
compatified open string case. Massive SGS, which is responsible for the
enhancement of massive gauge symmetry, are shown to exist at any higher
massive levels of the spectrum. A brief discussion is given in section IV.

\section{Chan-Paton Gauge States}

In this section, we discuss (zero-norm) gauge state of uncompatified open
string with Chan-Paton factor and its implication on on-shell symmetry and
Ward identity. For simplicity, we consider the oriented $U\left( N\right) $
case. The vertex operators of massless gauge state is

\begin{equation}
\theta ^{a}\lambda _{ij}^{a}k\cdot \partial xe^{ikx}
\end{equation}
where $\lambda \in U\left( N\right) ,i\in N,j\in \stackrel{\_}{N}$ and $a\in 
$ adjoint representation of $U\left( N\right) $. The on-shell conformal
deformation and $U\left( N\right) $ gauge symmetry to lowest order in the
weak background field approximation are $\left( \Box \theta ^{a}=0,\Box
\equiv \partial _{\mu }\partial ^{\mu }\right) $

\begin{equation}
\delta T=\lambda _{ij}^{a}\partial _{\mu }\theta ^{a}\partial x^{\mu }
\end{equation}
and

\begin{equation}
\delta A_{\mu }^{a}=\partial _{\mu }\theta ^{a}
\end{equation}
with T the energy momentum tensor and $A_{\mu }^{a}$ the massless gauge
field.

One can verify the corresponding Ward identity by calculating e.g., 1-vector
and 3-tachyons four point correlators. The amplitude is calculated to be

\begin{eqnarray}
T_{\mu }^{abcd} &=&\int \prod_{i=1}^{4}dx_{i}\left\langle
e^{ik_{1}x_{1}}\partial x_{\mu
}e^{ik_{2}x_{2}}e^{ik_{3}x_{3}}e^{ik_{4}x_{4}}\right\rangle T_{r}\left(
\lambda ^{a}\lambda ^{b}\lambda ^{c}\lambda ^{d}\right) \\
&=&\frac{\Gamma \left( -\frac{s}{2}-1\right) \Gamma \left( -\frac{t}{2}%
-1\right) }{\Gamma \left( \frac{u}{2}+1\right) }\left[ k_{3\mu }\left( \frac{%
s}{2}+1\right) -k_{1\mu }\left( \frac{t}{2}+1\right) \right] \times
T_{r}\left( \lambda ^{a}\lambda ^{b}\lambda ^{c}\lambda ^{d}\right) 
\nonumber
\end{eqnarray}
In equation (2.4), s , t and u are the usual Mandelstam variables. One can
then verify the Ward identity

\begin{equation}
\theta ^{b}k_{2}^{\mu }T_{\mu }^{abcd}=0
\end{equation}

We now discuss the massive gauge states. The vertex operator of type I
massive vector gauge state is

\begin{equation}
\theta _{\mu }^{a}\lambda _{ij}^{a}\left[ k\cdot \partial x\partial x^{\mu
}+\partial ^{2}x^{\mu }\right] e^{ikx}
\end{equation}
We note that the gauge state polarization contains both Einstein and
Yang-Mills indices. This is very similar to the 10d closed Heterotic string
case\cite{5}. The only difference is that in the Heterotic string, one could
have more than one Yang-Mills index. The on-shell conformal deformation and
the mixed Einstein-Yang-Mills-type symmetry to lowest order weak field
approximation are $\left( \left( \Box -2\right) \theta _{\mu }^{a}=\partial
\cdot \theta ^{a}=0\right) $

\begin{equation}
\delta T=\lambda _{ij}^{a}\partial _{\mu }\theta _{\nu }^{a}\partial x^{\mu
}\partial x^{\nu }+\lambda _{ij}^{a}\theta _{\mu }^{a}\partial ^{2}x^{\mu }
\end{equation}

and

\begin{equation}
\delta M_{\mu \nu }^{a}=\partial _{\mu }\theta _{\nu }^{a}+\partial _{\nu
}\theta _{\mu }^{a}
\end{equation}

One can also derive the corresponding massive Ward identity by calculating
the decay rate of one massive state to three tachyons. The most general
amplitude is calculated to be

\begin{equation}
A^{abcd}=\varepsilon ^{a}\varepsilon ^{c}\varepsilon ^{d}\left( \varepsilon
_{\mu \nu }^{b}T^{\mu \nu }+\varepsilon _{\mu }^{b}T^{\mu }\right) Tr\left(
\lambda ^{a}\lambda ^{b}\lambda ^{c}\lambda ^{d}\right)
\end{equation}

where

\begin{equation}
T^{\mu \nu }=\frac{\Gamma \left( -\frac{s}{2}-1\right) \Gamma \left( -\frac{t%
}{2}-1\right) }{\Gamma \left( \frac{u}{2}+2\right) }\left\{ \frac{s}{2}%
\left( \frac{s}{2}+1\right) k_{3}^{\mu }k_{3}^{\nu }+\frac{t}{2}\left( \frac{%
t}{2}+1\right) k_{1}^{\mu }k_{1}^{\nu }-2\left( \frac{s}{2}+1\right) \left( 
\frac{t}{2}+1\right) k_{1}^{\mu }k_{3}^{\nu }\right\}
\end{equation}

and

\begin{equation}
T^{\mu }=\frac{\Gamma \left( -\frac{s}{2}-1\right) \Gamma \left( -\frac{t}{2}%
-1\right) }{\Gamma \left( \frac{u}{2}+2\right) }\left\{ -k_{3}^{\mu }\frac{s%
}{2}\left( \frac{s}{2}+1\right) -k_{1}^{\mu }\frac{t}{2}\left( \frac{t}{2}%
+1\right) \right\}
\end{equation}
In equation (2.9) $\varepsilon ^{a}$ etc. are polarization of tachyons and $%
\left( \varepsilon _{\mu \nu }^{b},\varepsilon _{\mu }^{b}\right) $ is
polarization of the massive state. The above amplitude satisfies the
following ward identity

\begin{equation}
k_{\mu }\theta _{\nu }^{a}T^{\mu \nu }+\theta _{\mu }^{a}T^{\mu }=0
\end{equation}

Similar consideration can be applied to the following type II massive scalar
gauge state

\begin{equation}
\left[ \frac{1}{2}\alpha _{-1}\cdot \alpha _{-1}+\frac{5}{2}k\cdot \alpha
_{-2}+\frac{3}{2}\left( k\cdot \alpha _{-1}\right) ^{2}\right] \left|
k,l=0,i,j\right\rangle
\end{equation}

which corresponds to a {\em massive} $U\left( N\right) $ symmetry.

\section{Chan-Paton Soliton Gauge State on $R^{25}\otimes T^{1}$}

In this section, we discuss soliton gauge states on torus compatification of
bosonic open string. As is well known, the massless $U\left( N\right) $
gauge symmetry will be broken in general after compatification unless N
D-branes, in the T-dual picture, are coincident. We will see that when
D-branes are coincident, one has enhancement of (unwinding) zero-norm gauge
states and the massless $U\left( N\right) $ symmetry will be recovered.
These zero-norm gauge states can be considered as charges or symmetry
parameters of $U\left( N\right) $ group.

In the discussion of open string compatification, one needs to turn on the
wilson line or nonzero background gauge field in the compact direction. This
will effect the momentum in the compact direction, and the virasoro
operators become

\begin{eqnarray}
L_{0} &=&\frac{1}{2}\left( \frac{2\pi l-\theta _{j}+\theta _{i}}{2\pi R}%
\right) ^{2}+\frac{1}{2}\left( k^{\mu }\right) ^{2}+\sum_{n=1}^{\infty
}\left( \alpha _{-n}^{\mu }\alpha _{n}^{\mu }+\alpha _{-n}^{25}\alpha
_{n}^{25}\right) \\
L_{m} &=&\frac{1}{2}\sum_{-\infty }^{\infty }\stackrel{\rightharpoonup }{%
\alpha }_{m-n}\cdot \stackrel{\rightharpoonup }{\alpha }_{n}
\end{eqnarray}
Note that in equation (3.2), $\alpha _{0}^{25}\equiv p^{25}$ which also
appears in the first term in equation (3.1). k is the 25d momentum. $\theta
_{i},R$ are the gauge and space-time moduli respectively and $l$ is the
winding number in the compact direction. The spectrums of type I and type II
zero-norm gauge states become\cite{4}

\begin{equation}
M^{2}=\left( \frac{2\pi l-\theta _{j}+\theta _{i}}{2\pi R}\right) ^{2}+2I
\end{equation}

and

\begin{equation}
M^{2}=\left( \frac{2\pi l-\theta _{j}+\theta _{i}}{2\pi R}\right)
^{2}+2\left( I+1\right)
\end{equation}
where $I=\sum\limits_{n=1}^{\infty }\left( \alpha _{-n}^{\mu }\alpha
_{n}^{\mu }+\alpha _{-n}^{25}\alpha _{n}^{25}\right) $.

For the massless case $I=l=0$, one gets $N^{2}$ massless solution from
equation (3.3)

\begin{equation}
k_{\mu }\alpha _{-1}^{\mu }\left| k,l=0,i,j\right\rangle
\end{equation}
if all $\theta _{i}$ are equal, or in the T-dual picture when N D-branes are
coincident. These $N^{2}$ massless gauge states correspond to the charges of
massless $U\left( N\right) $ gauge symmetry. There is no type II massless
solution in equation (3.4).

We are now ready to discuss the interesting massive case. For $M^{2}=2$ and
general moduli $\left( R,\theta _{i}\right) $,

1. $I=1,l=0$, one gets two gauge states solutions from equation (3.3):

\begin{equation}
\left[ \left( \varepsilon \cdot \alpha _{-1}\right) \left( k\cdot \alpha
_{-1}\right) +\varepsilon \cdot \alpha _{-2}\right] \left|
k,l=0,i,i\right\rangle ,\text{ }\varepsilon \cdot k=0
\end{equation}

and

\begin{equation}
\left( k\cdot \alpha _{-1}\alpha _{-1}^{25}+\alpha _{-2}^{25}\right) \left|
k,l=0,i,i\right\rangle
\end{equation}

If all $\theta _{i}$ are equal, the $\left( i,i\right) $ is enhanced to $%
\left( i,j\right) $. Equation (3.7) implies a massive $U\left( N\right) $
symmetry with transformation parameter $\theta ^{a}$. Equation (3.6) implies
a massive Einstein-Yang-Mills-type symmetry with transformation parameter $%
\theta _{\mu }^{a}$

2. $I=0,\frac{2\pi l-\theta _{j}+\theta _{i}}{2\pi R}=\pm \sqrt{2}$, one
gets solution from equation (3.3)

\begin{equation}
\left( k\cdot \alpha _{-1}\pm \sqrt{2}\alpha _{-1}^{25}\right) \left|
k,l,i,j\right\rangle
\end{equation}
Now since $\left| \theta _{i}-\theta _{j}\right| <2\pi $, for any given $R$,
there is at most one solution of $\left( \left| l\right| ,\left| \theta
_{i}-\theta _{j}\right| \right) $. One is tempted to consider the case

\begin{equation}
\left( k\cdot \alpha _{-1}\pm \sqrt{2}\alpha _{-1}^{25}\right) \left|
k,l=\pm \sqrt{2}R,i,i\right\rangle
\end{equation}
That means in the moduli $\left( R=\sqrt{2}n,\theta _{i}\right) $ with $n\in
Z^{+}$, one has {\em soliton} gauge states which imply a {\em massive} $%
U\left( 1\right) ^{N}$ symmetry. If all $\theta _{i}$ are equal, the $\left(
i,i\right) $ is enhanced to $\left( i,j\right) $. Equation (3.9) implies a
massive $U\left( N\right) $ symmetry at the {\em discrete} values of moduli
point $R=\sqrt{2}n$. For example, in the T-dual picture, for $R=\sqrt{2}%
,l=\pm 2$, and if all D-branes are coincident, we have an enhanced massive $%
U\left( N\right) $ symmetry. This phenomenon is very different from the
massless case, where one gets enhanced $U\left( N\right) $ symmetry at {\em %
any} radius R when N D-branes are coincident.

We would like to point out that similar Einstein-Yang-Mills-type symmetry
was discovered before in the closed Heterotic string theory. There, however,
one could have more than one Yang-Mills indices on the transformation
parameters.

For the type II states with $M^{2}=2$ in equation (3.4), $I=l=0$. One gets
one more $U\left( N\right) $ gauge states

\begin{equation}
\left[ \frac{1}{2}\alpha _{-1}\cdot \alpha _{-1}+\frac{1}{2}\alpha
_{-1}^{25}\alpha _{-1}^{25}+\frac{5}{2}k\cdot \alpha _{-2}+\frac{3}{2}\left(
k\cdot \alpha _{-1}\right) ^{2}\right] \left| k,l=0,i,j\right\rangle
\end{equation}
if all $\theta _{i}$ are equal.

For the general mass level, choosing $I=0$ and $i,j$ in equation (3.2), we
have $l/R=\pm M$. For say $R=\sqrt{2}$ and $l=\pm \sqrt{2}M$, which implies

\begin{equation}
M^{2}=2n^{2},\text{ }n=0,1,2,\ldots
\end{equation}

So we have Chan-Paton soliton gauge states at any higher massive level of
the spectrum. Similar result was found in the closed string case.

\section{Conclusion}

Zero-norm gauge state solution in the old covariant quantization of string
theory is closely related to the BRST cohomology of the theory. Physically,
they correspond to charges of the symmetries\cite{6}. It is believed that
all space-time symmetry of string theory, including closed or open and
compatified or uncompatified ones, are due to the existence of (soliton)
gauge state in the spectrum. Similar consideration can be applied to the R-R
charges and D-branes. Presumably, there is no R-R zero-norm gauge states as
charges of R-R gauge fields in the type II string spectrum. How D-branes
carry the zero-norm gauge state charges to emit R-R fields is an interesting
question to study.

\section{Acknowledgments}

This research is supported by National Science Council of Taiwan, R.O.C.,
under grant number NSC 88-2112-M-009-011.

\end{document}